\documentclass{article}

\usepackage{graphicx}
\usepackage{psfig}
\usepackage{epsfig}
\usepackage[round]{natbib}

\title{Comparing the ensemble mean and the ensemble standard deviation as
inputs for probabilistic medium-range temperature forecasts}

\author{Stephen Jewson\footnote{\emph{Correspondence address}: RMS, 10 Eastcheap,
London, EC3M 1AJ, UK. Email: \texttt{x@stephenjewson.com}}\\
RMS, London, United Kingdom}
\begin{document}

\newcommand{\bx}[1]{\fbox{\begin{minipage}{15.8cm}#1\end{minipage}}}

\maketitle

\maketitle
\begin{abstract}
We ask the following question: what are the relative contributions of the
ensemble mean and the ensemble standard deviation to the skill of
a site-specific probabilistic temperature forecast? Is it the case
that most of the benefit of using an ensemble forecast to predict
temperatures comes from the ensemble mean, or from the ensemble
spread, or is the benefit derived equally from the two? The answer
is that one of the two is \emph{much} more useful than the other.
\end{abstract}

\section{Introduction}

We consider medium range probabilistic site-specific forecasts of
temperatures. There are a number of methods that can be used to produce such
forecasts. These methods differ in terms of a) the underlying
numerical forecast model, b) the predictors that are taken from
the forecast model and c) the statistical methods used to
transform these predictors into predictions. For instance, the
underlying model could be the ECMWF or the NCEP model (\emph{inter alia}); the
underlying predictors could be the ensemble mean and the
standard deviation or just the ensemble mean (\emph{inter alia}) and the statistical
transform used could be based on the rank histogram or spread
regression (\emph{inter alia}).

Much work has gone into trying to understand the sources of medium range
forecast skill, and to improve medium range forecasts. Some of this has
aimed at determining which of the various numerical models produce the best
predictors (point "a" above). There has also been some investigation
into which of the various statistical transforms are the most
appropriate (point "c" above). We will now address a third question but closely related question:
what is the relative importance of the different predictors that
can be derived from the numerical models (point "b" above). In
particular, we will consider the relative importance of the mean
and the standard deviation of ensemble forecast temperatures, 
when used to make a site-specific probabilistic forecast.
We consider \emph{only} the mean and the
standard deviation, rather than all the individual ensemble members,
since, for the dataset we will use, it has been shown by~\citet{jewson03i} that
there seems to be no useful information in the ensemble beyond the
mean and the spread.

The statistical models we use to address this question are all
derived from the spread regression  model of~\citet{jewsonbz03a}. 
This model takes the mean
and the standard deviation from an ensemble forecast and converts
them into a probabilistic forecast. This model is
particularly useful for addressing the question of relative
importance of the different predictors since it allows us to turn
each of them on and off rather easily.

We do not think that it is \emph{a priori} obvious which of the
ensemble mean and ensemble spread are more useful for making a
probabilistic forecast.
For instance,
one can imagine dynamical systems for which the expectation varies
very little, but for which the uncertainty varies a lot. Similarly
one can imagine dynamical systems for which the expectation varies
a lot, but for which the uncertainty varies little. In the first
of these it is likely that forecasts of the expectation are more
important, while in the latter it is likely that forecasts of the
uncertainty would be more important. We see, therefore, that our
question is, in part, a question about the dynamics of the
atmosphere. One can also imagine a particular forecast system that, for
whatever reason, does very well in predicting the expectation, but
very poorly in predicting the uncertainty, or vice versa. From
this we see that our question is also a question about specific
forecast systems.

In section~\ref{data} we describe the data we use for this study.
In section~\ref{models} we describe the methodology and the statistical
models we will use to address  this question. In
section~\ref{results} we present the results of our analysis and
in section~\ref{discussion} we discuss the implications. 

\section{Data}
\label{data}

We will base our analyses on one year of ensemble forecast data for the weather
station at London's Heathrow airport, WMO number 03772. The forecasts are predictions
of the daily average temperature, and the target days of the forecasts
run from 1st January 2002 to 31st December 2002. The forecast was produced
from the ECMWF model~\citep{molteniet96} and downscaled to the airport location using a simple
interpolation routine prior to our analysis. There are 51 members in the ensemble.
We will compare these forecasts to the quality controlled climate
values of daily average temperature for the same location as reported by the UKMO.

There is no guarantee that the forecast system was held constant throughout this period,
and as a result there is no guarantee that the forecasts are in any sense stationary,
quite apart from issues of seasonality. This is clearly far from ideal with respect to 
our attempts to build statistical interpretation models on past forecast data but is,
however, unavoidable: this is the data we have to work with.

Throughout this paper all equations and all values are in terms of double anomalies
(have had both the seasonal mean and
the seasonal standard deviation removed). 
Removing the seasonal standard deviation
removes most of the seasonality in the forecast error statistics, and partly justifies the use of
non-seasonal parameters in the statistical models for temperature that we propose.

\section{Models}
\label{models}

We will address the question of whether the useful information in
ensemble forecasts comes to a greater extent from the ensemble mean or from the
ensemble spread by considering the site-specific forecasts for
temperature at London Heathrow described above. 
Our forecasts are produced by four
\emph{ensemble interpretation models}, which is the name we give
to the statistical models that take the non-probabilistic predictors
from the ensemble forecasts and convert them into probability
density functions for the predicted temperatures. Such models are
essential if one wants to produce probability forecasts from
ensemble output. The models we choose are ideally suited for
answering the question at hand because they allow us to turn on
and off the various sources of information in the ensemble
very easily and see what impact that has on the skill
of the resulting forecast.

The simplest model we consider consists of the use of linear
regression between a \emph{single member} of the ensemble
and the observed temperature. The regression model serves to
correct biases, to adjust the amplitude of the
variability, and to
predict the uncertainty. 
We will write this model as:

\begin{equation}
    T_i \sim N(\alpha+\beta x_i, \gamma)
\end{equation}

where $T_i$ is the double anomaly temperature on day $i$ and
$x_i$ is the ensemble member on day $i$. 
The skill of this model gives an indication of the ability
of single integrations of the numerical model to predict the
distribution of future atmospheric states. It is a baseline that
the ensemble forecast should beat if it has any value.

Our second model is the same as the first, but the predictor is
the ensemble mean, rather than a single ensemble member. We write
this model as:

\begin{equation}
    T_i \sim N(\alpha+\beta m_i, \gamma)
\end{equation}

where $m_i$ is the ensemble mean on day $i$. This model should
hopefully perform better than the first model.

Our third model is an extension of the first model to include the
ensemble standard deviation as a predictor of the uncertainty:

\begin{equation}
    T_i \sim N(\alpha+\beta x_i, \gamma+\delta s_i)
\end{equation}

where $s_i$ is the ensemble standard deviation.

Finally our fourth model combines the second and the third models
by using the ensemble mean as a predictor for the temperature, and
the ensemble standard deviation as a predictor for the uncertainty
around the temperature. This is the spread regression model of~\citet{jewsonbz03a}.

\begin{equation}
    T_i \sim N(\alpha+\beta m_i, \gamma+\delta s_i)
\end{equation}

These models have been designed to allow us to answer the question
at hand. For instance, comparison of the results from model one
and model two, or between model three and model four, gives the
benefit of using the ensemble mean versus a single integration.
Comparison between model one and model three, or between model two
and model four, gives the benefit of using the ensemble spread versus
only using past forecast error statistics.

We fit all of these models using the standard maximum likelihood
method from classical statistics. More details of this method as
applied to this problem are given in~\citet{jewson03g}.

\subsection{Scoring the models}

In order to compare the results from our four models we need a
measure for how good the resulting probabilistic forecast is. 
To our knowledge only
two scores have been suggested in the literature for comparing
continuous probabilistic forecasts: the ''ignorance'' by~\citet{roulstons02}, and
the likelihood, by~\citet{jewsonbz03a}. It turns out that these two are, for this
case, more or less the same thing. We will therefore use this
score to compare our forecasts, in the form of the log-likelihood.

\section{Results}
\label{results}

\subsection{Parameter values}

\begin{figure}[!htb]
  \begin{center}
    \includegraphics{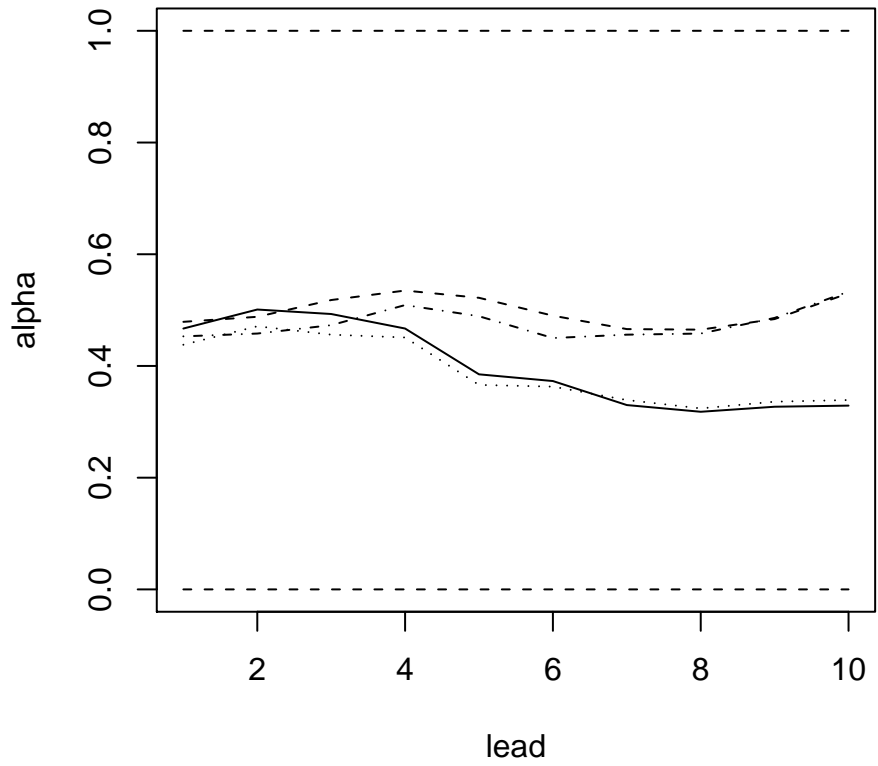}
  \end{center}
  \caption{
  The values of alpha for the four models described in the text.
  Model 1 (solid line), model 2 (dashed line), model 3 (dotted line) and model 4 (dot-dashed line).
          }
  \label{f:f1}
\end{figure}

We first look at the parameter values for our various statistical
models. Figure~\ref{f:f1} shows the values of alpha for the four
models. The two models that are based on the single ensemble
member have very similar values of alpha, while the two models
that are based on the ensemble mean also have similar values. The
values from the models based on the single ensemble member are
somewhat smaller.

\begin{figure}[!htb]
  \begin{center}
    \includegraphics{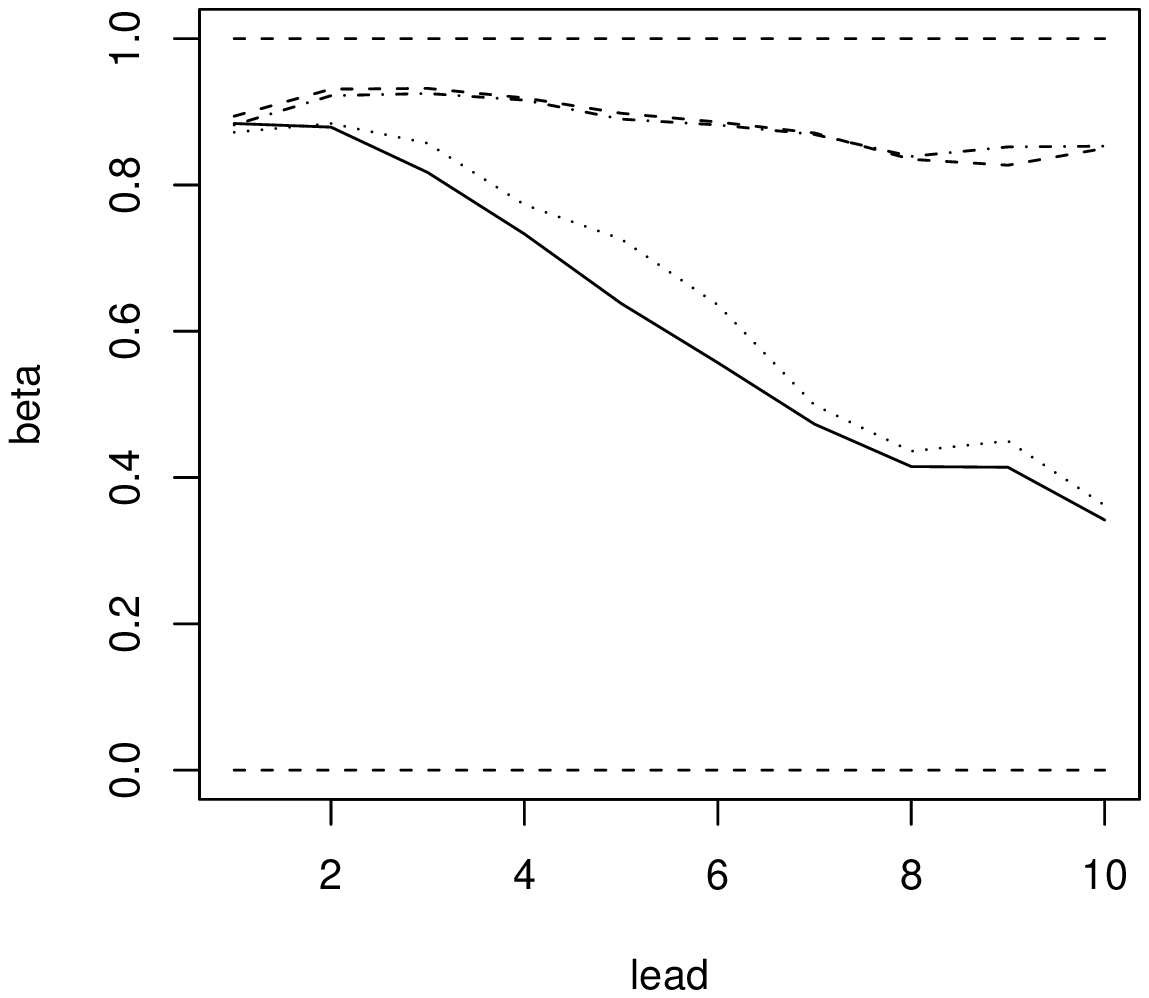}
  \end{center}
  \caption{
  The values of beta for the four models described in the text.
  Model 1 (solid line), model 2 (dashed line), model 3 (dotted line) and model 4 (dot-dashed line).
          }
  \label{f:f2}
\end{figure}

Figure~\ref{f:f2} shows the values of beta for the four models. The
values from the four models pair up in the same way as for the values of alpha,
with the values for the single model based forecasts being much lower
than the values for the ensemble mean based forecasts.

We can understand these values of beta as follows. We can
consider a single integration of the numerical model to consist of
the ensemble mean plus a noise term. The noise terms are at least
somewhat different for the different members of the ensemble.
Thus, in forming the ensemble mean the noise terms tend to
cancel out and the overall noise is smaller. 
When we regress a single member onto observed
temperatures, the regression coefficient is rather small because
of the large noise term. When we regress the ensemble mean onto
observed temperatures, the regression coefficient is much larger,
because the noise level is lower. The mathematics
of this effect have been given in detail by~\citet{jewsonz03a}.

The value of alpha can be understood in terms of the values of
beta. Alpha arises due to a combination of bias, and the beta
term. We would expect the bias to be the same for a single
ensemble member and the ensemble mean: the differences we see can
be explained as being due to the differences in the beta.

\begin{figure}[!htb]
  \begin{center}
    \includegraphics{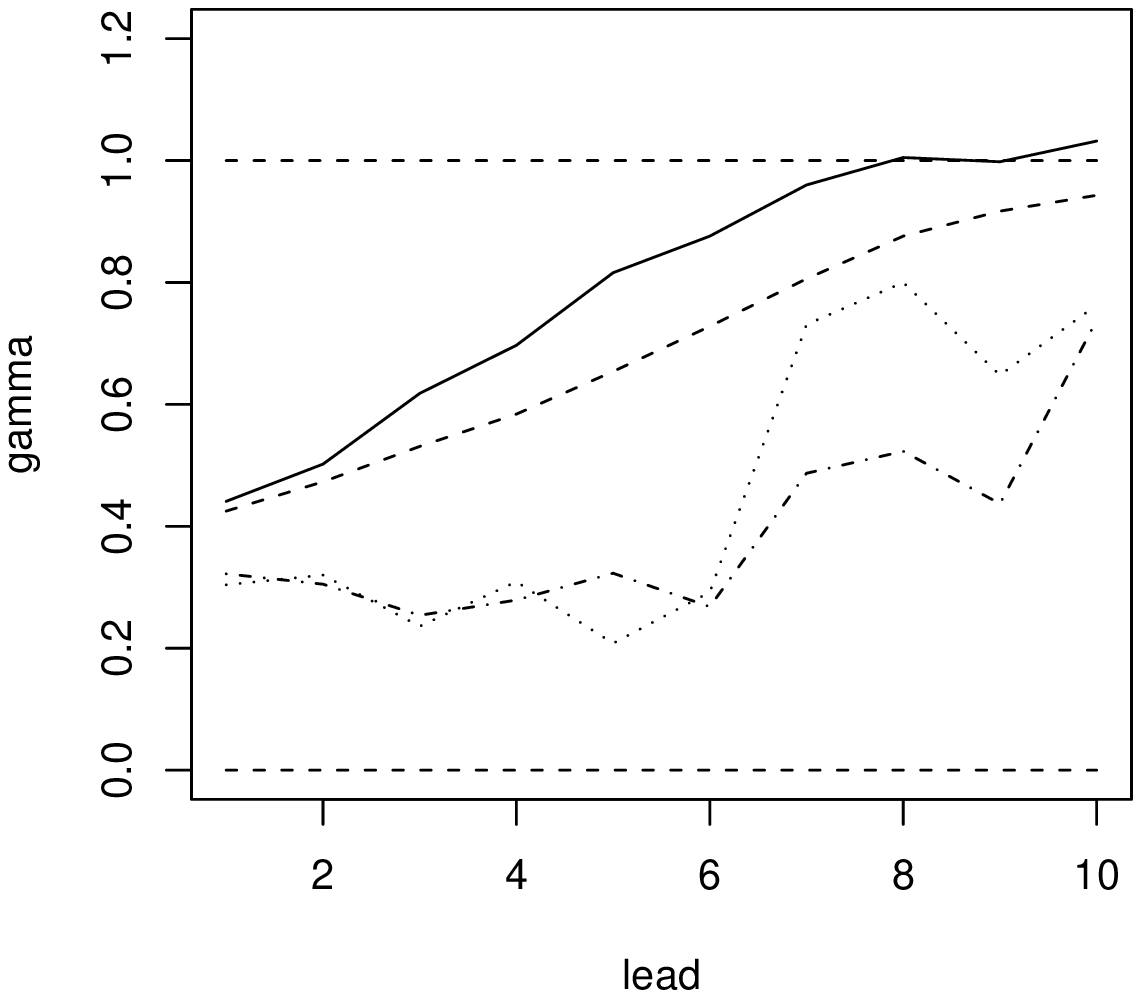}
  \end{center}
  \caption{
  The values of gamma for the four models described in the text.
  Model 1 (solid line), model 2 (dashed line), model 3 (dotted line) and model 4 (dot-dashed line).
          }
  \label{f:f3}
\end{figure}

\begin{figure}[!htb]
  \begin{center}
    \includegraphics{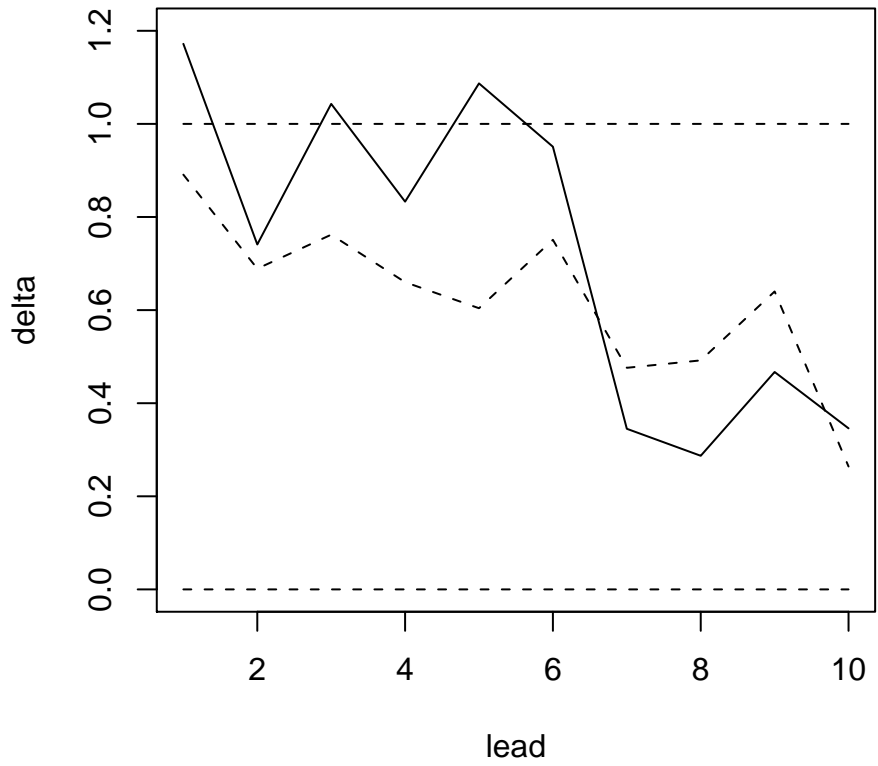}
  \end{center}
  \caption{
  The values of delta for two of the four models described in the text.
  Model 2 (solid line) and model 4 (dashed line).
           }
  \label{f:f4}
\end{figure}

Figure~\ref{f:f3} shows the values of gamma for the four models.
Because gamma plays a slightly different role in the different
models, there is no single physical interpretation of what it
means. In the two regression models gamma represents the
uncertainty: we see that the uncertainty is greater in the model
that only uses the single ensemble member, as would be expected.
In the spread regression models, gamma represents only a part of
the total uncertainty, and can only be interpreted in combination
with delta, which is shown in figure~\ref{f:f4}. 
Confidence intervals for the value of delta in the spread regression model are given by~\citet{jewsonbz03a},
and show that there is significant sampling uncertainty about these estimates.
This is not surprising since delta is related to the
second moment of the second moment of observed temperatures. This
sampling uncertainty is reflected in figure~\ref{f:f4} by the
jaggedness of the lines. Nevertheless we can see a general trend
towards lower values of delta at high leads for both models.

\subsection{Total Uncertainty}

\begin{figure}[!htb]
  \begin{center}
    \includegraphics{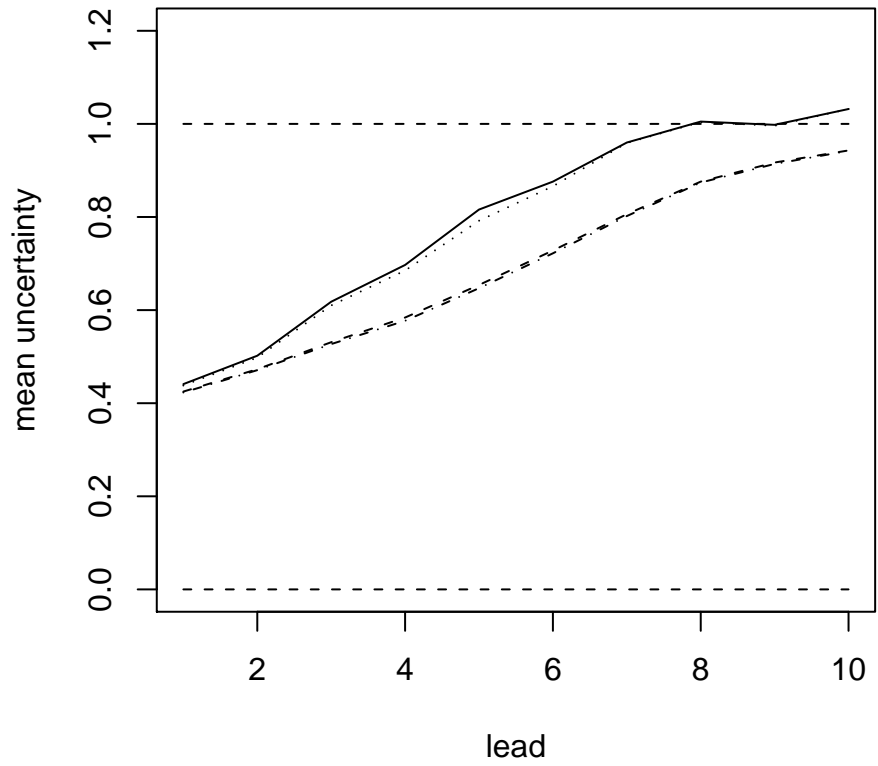}
  \end{center}
  \caption{
  The mean level of uncertainty predicted by the four models described in the text.
  Model 1 (solid line), model 2 (dashed line), model 3 (dotted line) and model 4 (dot-dashed line).
          }
  \label{f:f5}
\end{figure}

\begin{figure}[!htb]
  \begin{center}
    \includegraphics{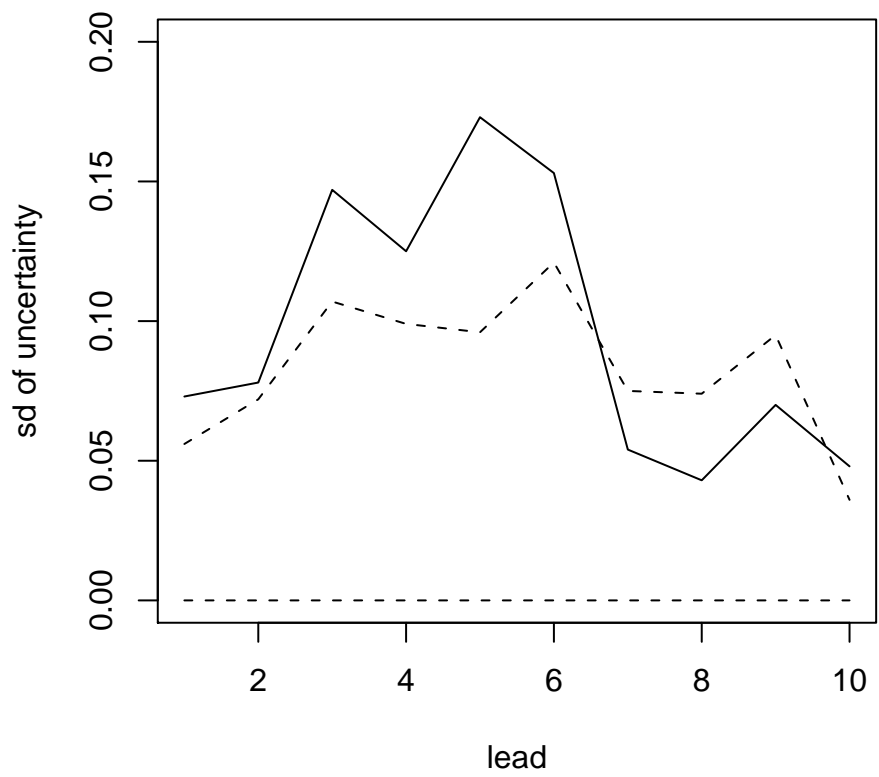}
  \end{center}
  \caption{
  The standard deviation of the level of uncertainty predicted by two of the four models described in the text.
  Model 2 (solid line) and model 4 (dashed line).
          }
  \label{f:f6}
\end{figure}

Comparing the gamma and delta parameter values for the four
statistical models is somewhat dissatisfactory since, as we have
seen, gamma does not have consistent interpretation across the
four models. It makes more sense to compare the total uncertainty
predicted by the four models. In figures~\ref{f:f5} and
figure~\ref{f:f6} we compare the mean and the standard deviation of
this predicted uncertainty respectively. 
In figure~\ref{f:f5} we see that the two
models based on the ensemble mean have much lower mean uncertainty
than the two models based on the single integration, as would be
expected. The ensemble mean gives significantly better forecasts
than individual ensemble members, especially at longer leads.
Figure~\ref{f:f6} shows the variability in the predicted uncertainty
for the two spread regression models. The two regression models,
both of which ignore the ensemble spread,
are not shown because they predict constant levels of uncertainty.
We see that the two spread regression models predict roughly the same levels
of variability in the uncertainty. There are some differences
between the two, but we suspect these differences are mostly due
to sampling error.

\subsection{Likelihood scores}

Finally we come to the results that answer the question that we
set out to address: what are the relative contributions of the
ensemble mean and the ensemble standard deviation to the skill of the final
forecast? We present the comparisons in terms of the negative of
the log-likehood, for which small values represent a good forecast. 
We compare the models in pairs to isolate individual effects, 
as described in section~\ref{models}.

\begin{figure}[!htb]
  \begin{center}
    \includegraphics{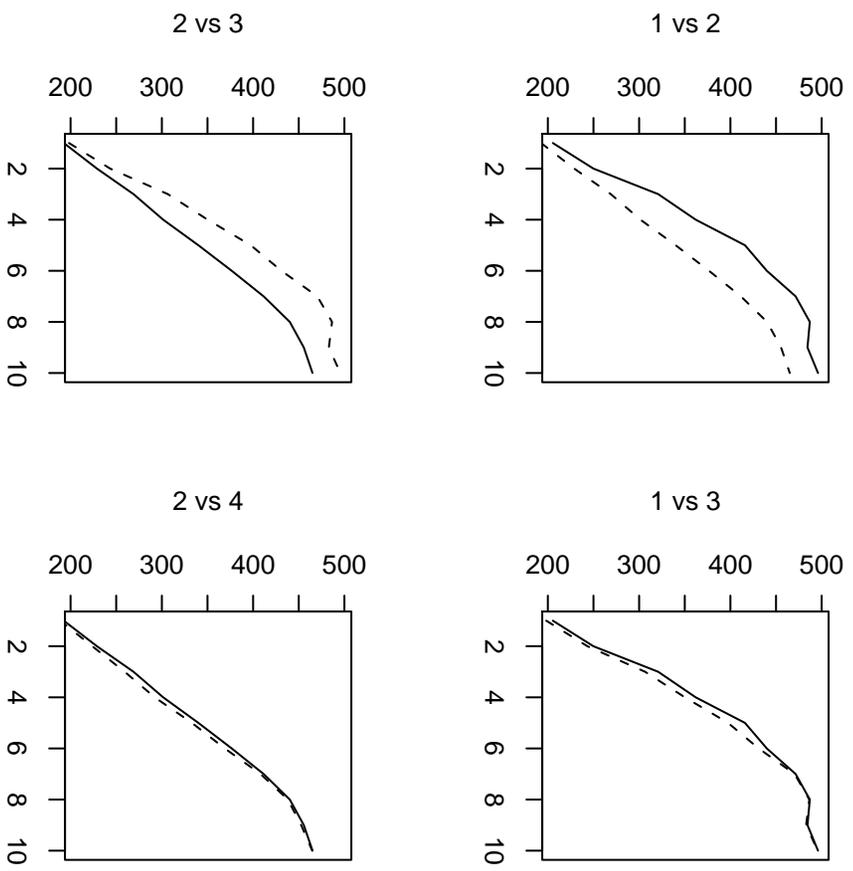}
  \end{center}
  \caption{
  Comparisons between the minus log-likelihood scores from the four models described in the text.
          }
  \label{f:f7}
\end{figure}

Figure~\ref{f:f7}, top left panel, shows a comparison between models
one and two. This comparison is designed to illustrate the benefit
of using the ensemble mean over using a single ensemble member. We
see a very large benefit at all lead times. For instance, the
ensemble mean based forecast at lead 6 is roughly as good as the
single member based forecast at lead 4. We are left in no doubt
that use of the ensemble mean is very beneficial for our
probabilistic forecast. The lower left panel shows a comparison
between models three and four. This comparison also shows the
benefit of using the ensemble mean over using a single ensemble
member, but now for the models that also use the ensemble spread.
The size of the benefit is roughly the same as before.

The right hand panels show the effects of using the ensemble
spread. The top right panel shows the benefit of using the
ensemble spread in the models that do not use the ensemble mean,
and the lower right panel shows the benefit of using the ensemble
spread in the models that do use the ensemble mean. The results
are roughly the same: in both cases we see only a very marginal
benefit from using the ensemble spread.

\begin{figure}[!htb]
  \begin{center}
    \includegraphics{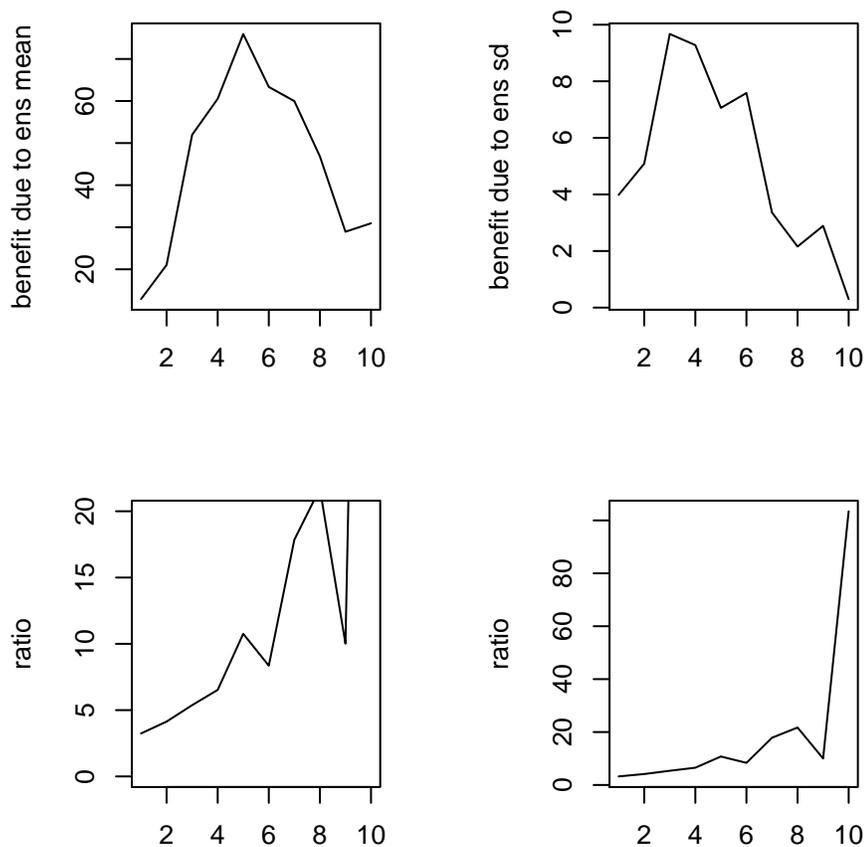}
  \end{center}
  \caption{
  Comparisons between the minus log-likelihood scores from the four models described in the text.
  The top left panel shows the differences in the log-likelihood between models 1 and 2, while the
  top right panel shows the differences between models 2 and 4.
  The lower panels show the ratio of these differences (with different vertical scales).
          }
  \label{f:f8}
\end{figure}

Figure~\ref{f:f8} now shows the differences that we see in the
benefit due to the ensemble mean and the ensemble spread. There
are a number of ways one could try and quantify this effect: we
choose to look at the numerical differences in the log-likelihood.
The top left hand panel shows the differences in the
log-likelihood caused by the use of the ensemble mean as a
predictor (from comparing models 1 and 2). The values lie between 10 and 70. 
The top right hand
panel shows the differences in the log-likelihood caused by the
use of the ensemble spread as a predictor (from comparing models 2 and 4). 
The values lie between
zero and 10. The lower panels show the ratios of these
log-likelihood differences, which quantify how much more useful the
ensemble mean is as a predictor than the ensemble spread. We see
the values for this ratio lie between four and 100, and increase
at longer lead times. The ensemble spread is relatively most
useful at the shortest lead times, while the ensemble mean is
relatively most useful at the longest lead times.

\section{Discussion}
\label{discussion}

We have investigated the relative contributions of the ensemble
mean and the ensemble standard deviation to the skill of
probabilistic temperature forecasts. This was done by using four
different statistical models to produce the probabilistic
forecasts. These models differ in terms of whether they use the
ensemble mean or a single ensemble member, and whether they use
the ensemble spread or predict uncertainty purely using past
forecast error statistics.

The results are very clear: we see that the ensemble mean is \emph{much}
more important than the ensemble spread. The difference is least
at short leads, where the ensemble mean is only about four times
as useful as the ensemble spread (using the particular
measure that we have chosen). At medium leads the ensemble
mean is between 10 and 20 times as useful as the ensemble spread,
and at the longest lead the ensemble mean is roughly 100 times
more useful than the ensemble spread.

There are a number of caveats to this study. In particular, we
have only used data from one station. Different results would
probably be obtained at other stations. We have also only used one
year of forecast data, and we have seen that this leads to rather
noisy estimates for some of our parameters. Our estimates would be
more accurate if we had more data, as long as that data were
stationary. Having said that, our main result, which is that the
ensemble mean is much more useful than the ensemble standard
deviation, seems to be so clear that we doubt whether it would
change even if we did repeat this analysis for a number of other
stations and with more data.

Finally, we note that have performed all of our analysis
in-sample. This is not ideal: if possible, out of sample results
are to be preferred. However, \citet{jewson03a} has shown that out of sample
one cannot detect \emph{any} beneficial effects of ensemble spread
whatsoever using this data-set, and so it would not be possible to
perform this study if we were to do it out of sample. By using an
in-sample analysis, and using statistical models with small numbers of
parameters to minimize overfitting and artificial skill, we have
been able to shed light on a question that could not be otherwise
addressed.

\newpage

\newpage

\newpage

\newpage

\newpage

\newpage

\newpage

\newpage

\bibliography{benefit}

\end{document}